\documentclass[a4paper,12pt]{article}

\usepackage[T2A]{fontenc}
\usepackage{amsmath,amssymb}
\usepackage{dsfont}
\usepackage{cite}
\usepackage[unicode,linktocpage=true,plainpages=false,pdfpagelabels=false]{hyperref}

\newcommand{\dd}{\partial}
\newcommand{\de}{\delta}
\newcommand{\m}{\mu}
\newcommand{\n}{\nu}
\newcommand{\ls}{\left(}
\newcommand{\rs}{\right)}
\newcommand{\La}{\Lambda}
\newcommand{\ka}{\varkappa}
\newcommand{\ta}{\tau}
\newcommand{\M}{{\mathcal M}}
\newcommand{\pp}{{\mathcal P}}
\newcommand{\zero}[1]{\mathop{#1}\limits^{\scriptscriptstyle (0)}{}\!}

\newcommand{\disn}[2]{$$\displaylines{\refstepcounter{equation}%
            \label{#1}\hskip 1em minus 1em #2\hfilneg}$$}
\newcommand{\nom}{\hfil\hskip 1em minus 1em (\theequation)}

\begin{document}

\title{Nontrivial isometric embeddings for flat spaces}
\author{S.~A.~Paston\thanks{E-mail: pastonsergey@gmail.com},
T.~I.~Zaitseva\thanks{E-mail: taisiiazaitseva@gmail.com}\\
{\it Saint Petersburg State University, Saint Petersburg, Russia}
}
\date{\vskip 15mm}
\maketitle

\begin{abstract}
Nontrivial isometric embeddings for flat metrics (i.e., those which are not just planes in the ambient space) can serve as useful tools in the description of gravity in the embedding gravity approach. Such embeddings can additionally be required to have the same symmetry as the metric. On the other hand, it is possible to require the embedding to be unfolded so that the surface in the ambient space would occupy the subspace of the maximum possible dimension. In the weak gravitational field limit, such a requirement together with a large enough dimension of the ambient space makes embedding gravity equivalent to General
Relativity, while at lower dimensions it guarantees the linearizability of the equations of motion. We discuss symmetric embeddings for the metrics of flat Euclidean three-dimensional space and Minkowski space. We propose the method of sequential surface deformations for the construction of unfolded embeddings. We use it to construct such embeddings of flat Euclidean three-dimensional space and Minkowski
space, which can be used to analyze the equations of motion of embedding gravity.
\end{abstract}

\newpage

\section{Introduction}
According to the Janet-Cartan-Friedman (JCF) theorem \cite{fridman61}, an arbitrary $n$-dimensional pseudo-Riemannian space can be locally isometrically embedded into the ambient flat space of dimension $N\geqslant n(n+1)/2$ with suitable signature.
By isometric embedding we mean the surface described by the embedding function $y^a(x^\m)$ in the ambient space for which the induced metric
\begin{equation}\label{first_formula}
g_{\mu \nu} = \big( \partial _\mu y^a  \big)\big(\partial _\nu y^b  \big) \eta _{ab}
\end{equation}
coincides with the metric of the original pseudo-Riemannian space.
Hereinafter, Greek indices $\m,\n,\ldots$ run over $n$ values; Latin indices $a,b,\ldots$ run over $N$ values;
$\eta_{ab}$ is the flat metric of the ambient pseudo-Euclidean space.
For specific pseudo-Riemannian spaces the required dimension of the ambient space can decrease.
In particular, this happens if a space has a sufficiently large number of symmetries \cite{schmutzer}.
The difference $N-n$ is called the \emph{class of the embedding}.
We emphasize that the JCF theorem consider only \emph{local} embeddings, and when passing to \emph{global} embeddings, the required dimension of the ambient space increases sharply \cite{kobno}.
However, in specific cases, even for global embeddings, the class of the embedding can be smaller, see, for example, \cite{statja40, statja56}.

The interest in explicit isometric embeddings of physically meaningful pseudo-Riemannian spaces is due to several reasons.
First of all, it provides an opportunity to understand the geometric structure of space-time better since this structure manifests itself in the presence of an explicit embedding.
That is why a great interest in the construction of explicit embeddings has been shown by researchers in the case of various black hole metrics.
For the Schwarzschild metric, the first embedding \cite{kasner3} was proposed just 5 years after its discovery.
The only global embedding \cite{frons} turns out to be closely related to the maximum analytic extension of the Schwarzschild metric by the Kruskal-Shekeres coordinates (see the note at the end of \cite{frons}).
In general, a lot of works have been devoted to the construction of explicit embeddings of various black holes, including charged and rotating ones; see, for example, the links in \cite{statja56}.
Other physically interesting metrics include various cosmological models, e.g. Friedmann-Robertson-Walker metric that describe the expanding universe.
Explicit embedding of this metric \cite{robertson1933} was also found a very long time ago.
It should be noted that the problem of finding an embedding for a given metric usually has more than one solution since when solving differential equations, arbitrary parameters (numbers or functions) arise. A smooth surface deformation that does not alter the induced metric is called isometric bending; see, for example, the discussion of this question in \cite{statja64} and references therein.

Other motivations for considering isometric embeddings include their use in the  classification of exact solutions of the Einstein equations \cite{schmutzer}, as well as in the calculation of Hawking temperature of spacetimes with a horizon (see, for example, references in \cite{statja36}).
However, from the point of view of describing gravity, the main motivation is the possibility of obtaining a modified theory of
gravity by variable substitution \eqref{first_formula} in General Relativity (GR) action with matter contribution $\mathcal{L}_m$
\begin{equation}\label{ein-hilb-action}
S = \int\!  d^4 x \sqrt{-g} \ls - \frac{1}{2\ka}R + \mathcal{L}_m\rs,
\end{equation}
where $n=4$.
After such a substitution the theory might change (additional solutions appear, see \cite{statja60} for a discussion of gravity modifications  resulting from differential transformations of field variables). It happens even if the number of new variables $y^a(x)$, which is equal to the number of ambient space dimensions,
%corresponding to the JCF theorem,
corresponds to the JCF theorem value $N=10$ and therefore
does not differ from the number of the old metric variables $g_{\m\n}(x)$.

This string-inspired approach  was first proposed in \cite{regge} and was  subsequently studied in a number of works \cite{deser,pavsic85let,maia89,davkar,statja18,estabrook2009,statja25,faddeev} under the names like embedding theory, geodetic brane gravity and embedding gravity.
From variation of the action \eqref{ein-hilb-action} with respect to the independent variable $y^a(x)$ the Regge-Teitelboim (RT) equations \cite{regge} arise:
\begin{equation}\label{rt_equations}
%D_{\mu}\big((G^{\mu \nu} - \ka\, T^{\mu \nu}) \dd_\nu y^a\big) = 0,
(G^{\mu \nu} - \ka\, T^{\mu \nu})b_{\m\n}^a
% D_{\mu}\dd_\nu y^a
= 0.
\end{equation}
Here $G^{\mu \nu}$ is the Einstein tensor, $T^{\mu \nu}$ is the energy-momentum tensor of matter,
and
\disn{p2}{
b_{\m\n}^a=D_{\mu}\dd_\nu y^a,
\nom}
where $D_\mu$ is the covariant derivative,
is called the second fundamental form of the surface described by the embedding function $y^a(x)$ (for example, see \cite{goenner}).

It is easy to see that the RT equations \eqref{rt_equations} are more general than the Einstein equations: any solution to the Einstein equations is a solution to the RT equations, but not vice versa.
In addition to Einstein's solutions, there are so-called "extra" solutions.
As a result, the theory is not equivalent to GR.
In \cite{regge} this was regarded as a problem since the goal of this paper was to obtain an equivalent reformulation of General Relativity (mainly in the hope of advancing in the quantization of the theory) and not a transition to a more general theory.
For this reason, additional conditions called Einstein constraints were imposed in \cite{regge} and several subsequent papers.
However, at present, in connection with the modern cosmological mystery of dark matter and dark energy, the transition to modified theories of gravity which are more general than GR becomes more attractive. Along this path one can try to interpret \emph{extra} solutions as effects associated with dark matter or dark energy within the framework of GR.
The  mimetic gravity \cite{mukhanov,Golovnev201439,mimetic-review17} is the most famous approach of such kind, but the embedding theory approach is also possible \cite{davids97,davids01,statja51,statja68}.

In the analysis of solutions of the equations \eqref{rt_equations} as the equations of modified gravity one usually starts from weak field limit when the metric $g_{\m\n}$ is close to the flat metric $\eta_{\m\n}$.
Such a problem corresponds, for example, to the description of observations on the scale of the solar system or a galaxy.
Then one should determine the form of explicit embedding of the flat metric which can serve as the background solution $\zero{y}^a(x)$ in order to look for solutions of the equations \eqref{rt_equations} corresponding to a weak gravitational field in the form
\begin{equation}\label{pp1}
y^a(x) = \zero{y}^a(x) + \delta y^a(x).
\end{equation}

The simplest option is to select a plane as the background surface:
\disn{p1}{
\zero{y}^a(x)=\de^a_\m x^\m.
\nom}
However, as noted in \cite{deser}, when such a background is used, the equations of the embedding theory \eqref{rt_equations} turn out to be non-linearizable (non-linear with respect to variation $\delta y^a$).
The easiest way to see it is to write down the Einstein tensor in the form (see, for example, \cite{statja18})
\begin{equation}
G^{\mu \nu} = \frac{1}{2} g_{\xi \zeta} E^{\mu \xi \alpha \beta} E^{\nu \zeta \gamma \delta} b^b_{\alpha \gamma}\eta_{bc} b^c_{\beta \delta}
\end{equation}
where $E^{\mu \xi \alpha \beta} = \varepsilon ^{\mu \xi \alpha \beta} /\sqrt{|g|}$ is the covariant unit antisymmetric tensor.
Then the RT equations \eqref{rt_equations} take the form
\begin{equation}\label{rt_upgraded}
g_{\xi \zeta} E^{\mu \xi \alpha \beta} E^{\nu \zeta \gamma \delta} b^b_{\alpha \gamma} \eta_{bc} b^c_{\beta \delta} b^a _{\mu \nu} =2\ka\, T^{\mu \nu} b_{\m\n}^a
\end{equation}
where the left-hand side is cubic in $b^a _{\mu \nu}$, which according to \eqref{p2} is linear in the small variation $\delta y^a$ when using the background \eqref{p1}.

Since for a weak gravitational field the principle of superposition must be satisfied, the non-linearity of the equation describing it seems unnatural.
On the other hand, the formula \eqref{p1} is far from the only possible choice of the background embedding function for a weak gravitational field.
As such, one can use any embedding of the flat metric $\eta_{\m\n}$ which can be quite nontrivial.
This paper is devoted to the problem of constructing such nontrivial embeddings.

In the next section we discuss possible additional requirements that can be imposed on the sought nontrivial embeddings.
The concept of "unfolded" embedding  introduced here turns out to be close to the concept of "free" embedding introduced in \cite{bustamante}; section \ref{razdbust} is devoted to their comparison.
In section \ref{razdsym}, we discuss the construction of symmetric embeddings for flat metrics.
We propose a nontrivial explicit symmetric embedding for the Minkowski metric.
It turns out to be non-unfolded.
As a result, if it is used as the background in the analysis of the RT equations solutions, linearization is only partial.
In section \ref{razdrazv} we propose a method for constructing unfolded embeddings for flat metrics; we construct such explicit embeddings for flat 3-dimensional Euclidean space and Minkowski space.

\section{Symmetrical and unfolded embeddings}\label{razdtreb}
When looking for nontrivial embeddings of a flat metric, some additional requirements can be imposed.
It seems natural enough to assume that the surface resulting from the embedding must have the same symmetry as the embedded metric.
For example, the metric $\eta_{\m\n}$ of Minkowski space $\mathds{R} ^{1,3}$  is symmetric with respect to the Poincare group $SO(1,3) \ltimes T^4$, and one can require the constructed surface to have the same symmetry.
We say that a surface $\M$ is \textit{symmetric} with respect to the group $G$ if $\M$ transforms into itself under the action of some  subgroup of the group of motions $\pp$ of the flat ambient space $\mathds{R}^{n_+,n_-}$ when this subgroup is isomorphic to $G$.
In \cite{statja27} a method for constructing explicit embeddings with a given symmetry was proposed;
it has recently been developed further in \cite{statja70}.
The idea is to consider all possible $N$-dimensional representations for a given symmetry group with the subsequent selection of those that have the form of transformations of group of motions $\pp$ of the flat ambient space and lead to the correct surface dimension; see details in \cite{statja27}.
We discuss the construction of symmetric embeddings for flat metrics in section \ref{razdsym}.

Another interesting requirement that can be imposed when searching for a background embedding function for a weak gravitational field is, in a sense, the maximal non-degeneracy for the second fundamental form \eqref{p2}.
Note that this object is symmetric with respect to the permutation of the indices $\m$ and $\n$, so this pair of indices can be replaced with a multi-index running through $n(n+1)/2$ values.
On the other hand, for the second fundamental form $b_{\m\n}^a$ the following identity holds (see, for example, \cite{goenner}):
\disn{p3}{
b_{\m\n}^a \dd_\beta y_a=0,
\nom}
which means that the index $a$ of $b_{\m\n}^a$ is transverse.
After the introduction of some basis in the space orthogonal to the surface at a given point,  it is possible  to replace the index $a$ of the object $b_{\m\n}^a$ with a new index running through $N-n$ values.
As a result, this value can be interpreted as a matrix of size $n(n+1)/2$ by $(N-n)$.
By the maximal non-degeneracy of $b_{\m\n}^a$ we mean that the rank of such a matrix is maximal, in other words, it coincides with its minimal dimension.
A surface is called unfolded if its second fundamental form satisfies this requirement at all points except, perhaps, a set of zero measure.
Violation of this requirement at some point geometrically means that in the neighborhood of this point the surface lies in some subspace of the ambient space the dimension of which is less than possible, i.e. one can additionally "unfold" the surface.

If, for example, for a 4-dimensional surface (i.e., $n=4$) we take the dimension of the ambient space $N=14$, then $b_{\m\n}^a$ turns out to be a square $10 \times 10$ matrix in the indicated sense.
If, in this case, the unfolded embedding of the flat metric is chosen as the background in the decomposition \eqref{pp1}, then within the framework of the perturbation theory $b_{\m\n}^a$ \eqref{p2} can be removed from the RT equation \eqref{rt_equations} as a non-singular matrix that is a factor in a homogeneous equation.
As a result, in this case, the RT equations are completely equivalent to the Einstein equations.
Thus, for $N=14$ and an unfolded embedding as a background the embedding theory becomes exactly equivalent to GR in the weak field limit.
This creates relevance for the problem of constructing an explicit unfolded embedding of the flat metric with $N=14$.
We propose a way to construct such an embedding in section \ref{razdrazv}.

In the most frequently discussed $N=10$ case (which corresponds to the minimal dimension by the JCF theorem) $b_{\m\n}^a$ is a non-square matrix and it cannot be removed from the equation, which means that the theory contains \emph{extra} solutions.
In this case, the RT equations \eqref{rt_equations} can be rewritten \cite{pavsic85let} as the system of the equations
 \disn{r3.1}{
G^{\m\n}=\ka \ls T^{\m\n}+\ta^{\m\n}\rs,
\nom}\vskip -2em
 \disn{r3.2}{
\ta^{\m\n} b_{\m\n}^a=0.
\nom}
The first of them is the Einstein equation with an additional contribution $\ta^{\m\n}$ of some fictitious matter (the properties of which can be compared with the known properties of dark matter or energy), and the second plays the role of equations of motion of this matter, see details in \cite{statja68}.
This allows us to consider embedding theory as a possible way to explain the mystery of dark matter. The latter turns out to be a purely gravitational effect that arises when considering the solutions of the embedding theory equations from the Einsteinian point of view.
When analyzing the properties of such matter in the non-relativistic limit, the embedding $y^a(x^\m)$ in section $x^0=const$ is an unfolded embedding of the flat three-dimensional Euclidean metric into the 9-dimensional ambient space \cite{statja68}.
In general, the
zeroth-order
embedding function $\zero{y}^a(x)$ of a four-dimensional surface
% to a zero approximation
is an unfolded embedding of the metric of the Minkowski space into the 10-dimensional ambient space.
To continue research in this direction, it is necessary to have an explicit form of such embeddings, and in this paper we find examples of these embeddings in section~\ref{razdrazv}.

\section{The relation between unfolded and free embeddings}\label{razdbust}
In the previous Section we introduced the notion of \emph{unfolded} embedding, which is convenient when discussing \emph{extra} solutions of the RT equations. It is closely related to the classification of embeddings introduced in \cite{bustamante}, which includes so-called \emph{free} embeddings, \emph{$q$-free} embeddings and \emph{spatially-free} embeddings.
Let's discuss their relation.

Let us consider an embedding function of an $n$-dimensional surface $y^a(x)$.
Let us investigate how its small variation $\delta y^a(x)$ affects the induced metric $g_{\mu \nu}$ in the lowest order.
An arbitrary variation $\delta y^a(x)$ can be decomposed into longitudinal and transverse contributions:
\disn{p4}{
\delta y^a=\xi^\m\dd_\m y^a+\delta y^a _{\perp}, \qquad \delta y^a _{\perp}\dd_\m y_a=0.
\nom}
Then from  \eqref{first_formula} we find that
\begin{equation}\label{urav_bustamante}
\delta g_{\mu \nu} =  (\partial _\mu y_a)(\partial _\nu \delta y^a)+(\partial _\n y_a)(\partial _\m \delta y^a) =
D _\nu\xi_\m+D _\m\xi_\n  -2 b^a _{\mu \nu} \delta y_{a \perp},
\end{equation}
where \eqref{p2} is used, as well as the properties of the covariant derivative in the embedding theory formalism, see details, for example, in \cite{statja18}.

In \cite{bustamante} an embedding for a $n$-dimensional metric is said to be "free" if the system of ${n(n+1)}/{2}$ equations(\ref{urav_bustamante}) can be solved in the transverse variations $\delta y^a_{\perp}$ for any $\delta g_{\mu \nu}$ and $\xi^\mu$.
Since $\delta y^a_{\perp}$ has $N-n$ independent components, a free embedding satisfies the relation
\disn{p5}{
N-n \geqslant \frac{n(n+1)}{2}\quad\Rightarrow\quad
N \geqslant \frac{n(n+3)}{2}.
\nom}
In turn, an embedding is "$q$-free" if $q$ out of ${n(n+1)}/{2}$ equations can be solved in the transverse variations $\delta y^a_{\perp}$ for any $\xi ^\mu$, and the ${n(n+1)}/{2} - q$ remaining equations are constraints on $\xi^\mu$.
It is easy to see that $q$ is the rank of the matrix constructed from the second fundamental form $b_{\m\n}^a$ in the manner described after \eqref{p3}.
The definition of $q$-free embedding implies that
\disn{p5.1}{
N-n\geqslant q, \qquad
q \leqslant \frac{n(n+1)}{2}, \qquad
n \geqslant \frac{n(n+1)}{2} - q.
\nom}
As a result, for a $q$-free embedding the following relations hold:
\disn{p6}{
\frac{n(n-1)}{2} \leqslant q \leqslant \frac{n(n+1)}{2}, \qquad
N \geqslant q+n,
\nom}
moreover, for $q = {n(n+1)}/{2}$ a $q$-free embedding is a free embedding.

From the inequalities \eqref{p5.1} it follows that for $q$-free embeddings (and, in particular, for free embeddings) $N \geqslant {n(n+1)}/{2}$.
This corresponds to the fact that within the statement of the problem (\ref{urav_bustamante}) we consider the perturbation of the metric $\delta g_{\mu \nu}$ to be arbitrary, and to embed an arbitrary metric, according to the JCF theorem, this restriction on the number of dimensions of the ambient space $N$ must be satisfied.
Note that for a four-dimensional surface, i.e. for $n=4$, $q$ lies in the interval $6 \leqslant q \leqslant 10$, and
a free embedding is $10$-free with $N \geqslant 14$.

Also in \cite{bustamante} the concept of "spatially-free" embedding is introduced.
A four-dimensional embedding is called spatially-free if one can choose a coordinate system $(x^0,x^i)$ such that
%components of the second fundamental form \eqref{p2}
 $6$ vector fields $b_{ij}^a$ (for $i,j=1,2,3$) are linearly independent.
Note that the type of a 10-dimensional unfolded embedding of the 4-dimensional Minkowski space metric described at the end of section \ref{razdtreb} (and used in the work \cite{statja68}) is spatially free.

According to the definition (see after \eqref{p3}) of an unfolded embedding introduced above, the matrix constructed from the second fundamental form $b_{\m\n}^a$ has the maximal possible rank.
It is easy to see that for a free embedding, for which the requirement \eqref{p5} is satisfied, the rank must also be subject to this condition, therefore, for
\disn{p6.1}{
N \geqslant {n(n+3)}/{2}
\nom}
the concepts of unfolded embedding and free embedding coincide.
For smaller values of the dimensions of the ambient space $N$, namely, in the range ${n(n+1)}/{2}\leqslant N \leqslant {n(n+3)}/{2}$, any unfolded embedding is a $(N-n)$-free embedding, but generally speaking, not vice versa.
As mentioned above, for even smaller values of $N < {n(n+1)}/{2}$ the concept of $q$-free embedding is no longer applicable while the concept of unfolded embedding can still be used.

The concept of unfolded embedding is very important in the perturbative analysis of solutions of RT equations \eqref{rt_equations} based on the decomposition \eqref{pp1}.
If one takes an unfolded embedding as a background, then the equations acquire interesting properties.
For $N \geqslant {n(n+3)}/{2}$, when the embedding is also free, the factor $b_{\m\n}^a$ \eqref{p2} can be removed since for an unfolded embedding all vectors of the ambient space $b_{\m\n}^a$ are linearly independent for $\m\geqslant\n$.
As a result, the RT equations reduce to the Einstein equations, and there will be no extra solutions in this case.

For smaller values of $N$, the usage of the unfolded embedding as a background guarantees the linearizability of the RT equations in the weak field limit.
To see this, note that an arbitrary deformation of the surface can be defined by choosing the deformation $\de y^a$ transverse to the surface in \eqref{pp1} so that an arbitrary deformation is defined by $N-n$ functions.
In the case of a weak gravitational field in zeroth-order of perturbation theory the metric must be flat, therefore $\zero{G}^{\m\n}=0$.
Thus, on the left-hand side of the RT equation \eqref{rt_upgraded} the factor $b_{\m\n}^a$ should be taken to a zeroth-order.
Linearization of the equations means that all $N-n$ functions defining an arbitrary deformation are solutions of linear equations, and for this the presence of $N-n$ independent equations is necessary.
This is exactly what happens if there are $N-n$ linearly independent vectors among the ambient space vectors $\zero{b}_{\m\n}^a$ for different $\m,\n$, which means that the embedding is unfolded for $N < {n(n+3)}/{2}$.

\section{Explicit symmetric embeddings of flat metrics}\label{razdsym}
Firstly, we will consider the explicit symmetric embedding of the flat Euclidian 3-dimensional (i.e. $n=3$) metric which has the $SO(3) \ltimes  T^3$ symmetry.
Such an embedding can be constructed by the method proposed in \cite{statja27} which was discussed at the beginning of section \ref{razdtreb}.
This method is implemented in \cite{statja29} for $N=5$ as an illustration of its application to the construction of  embedding for a spatially flat FRW model since the sections $x^0=\text{const}$ of spatially flat FRW spacetime have a flat Euclidean 3-dimensional metric.
However, there is a simple alternative possibility of obtaining the same embedding.

Consider the 4-dimensional surface that is a hyperboloid of radius $R$ in the space $\mathds{R}^{1,4}$ of signature $\{-++++\}$:
\disn{p7}{
y^{0} =\frac{1}{4t} (R^2 - x^i x^i)-t,\qquad
y^{1} =\frac{1}{4t} (R^2 - x^i x^i)+t,\qquad
y^{i} =x^i,
\nom}
where $i=1,2,3$ and $t,x^i$ are the coordinates on the surface.
It is easy to see that $y^a y^b \eta_{ab}=R^2$, i.e. it is indeed a hyperboloid.
Note that \eqref{p7} is an embedding of de Sitter space and the surface defined by this embedding is symmetric with respect to the group $SO(1,4)$.
If we take the section $y^0=\bar y^0=const$ of the given surface at large values of $\bar y^0$,
then the resulting 3-dimensional surface is a sphere of large radius which is locally indistinguishable from the 3-dimensional plane.
Thus, in the limit, the symmetry of the section is the group $SO(3)\ltimes T^3$.
If we make a hyperbolic rotation in the plane $y^0,y^1$ (which is an element of the group $ SO(1,4)$) simultaneously with increasing $\bar y^0$,
then in the limit $\bar y^0\to\infty$ the section $y^0=\bar y^0$ turns into the section
\disn{p8a1}{
y^1-y^0=2\bar t,
\nom}
where $\bar t$ is some constant, finite in the limit.
Therefore we see that the 3-dimensional surface resulting from the section \eqref{p8a1} has the desired symmetry $SO(3)\ltimes T^3$.
After an insignificant general translation of this surface its embedding function can be written as
\disn{p8a2}{
y^{0} =y^{1} =-\frac{1}{4t_0} x^i x^i,\qquad
y^{i} =x^i,
\nom}
corresponding to the symmetric 5-dimensional embedding of the flat Euclidean 3-dimensional metric (the fact that the resulting surface has the flat Euclidean induced metric can easily be checked directly by the formula \eqref{first_formula}).

We will use an analogous method to construct an embedding of the 4-dimensional Minkowski space metric with the Poincare group symmetry $SO(1,3)\ltimes T^4$, i.e. $n=4$.
Consider the 5-dimensional surface that is a hyperboloid of radius $R$ in the space $\mathds{R}^{2,4}$ of signature $\{+----+\}$:
\disn{p8}{
y^{\m} =x^\m,\qquad
y^{4} =\frac{1}{4t} (R^2 - x^\m x^\n\eta_{\m\n})-t,\qquad
y^{5} =\frac{1}{4t} (R^2 - x^\m x^\n\eta_{\m\n})+t,
\nom}
where $\m=0,1,2,3$;
$t,x^i$ are the coordinates on the surface,
and $\eta_{\m\n}$ is the Minkowski metric of signature $\{+---\}$.
It is easy to check that
\disn{p8.1}{
y^a y^b \eta_{ab}=R^2,
\nom}
i.e. it is indeed a hyperboloid.
Note that \eqref{p8} is an embedding of 5-dimensional anti-de Sitter space $AdS_5$, and the surface is symmetric with respect to the group $SO(2,4)$.
Take a section of the hyperboloid \eqref{p8} by the plane
\disn{p8.2}{
y^{4}=\bar y^{4}=const.
\nom}
The resulting 4-dimensional surface, according to \eqref{p8.1}, is described by the equation
\disn{p8.3}{
y^\m y^\m \eta_{ab}+ \big({y^5} \big)^2=R^2+\big( {\bar y{}^4} \big) ^2,
\nom}
hence, it is a 4-dimensional hyperboloid of radius $\sqrt{R^2+{\bar y{}^4}^2}$.
For $\bar y^4\to\infty$ it is locally indistinguishable from the 4-dimensional plane of signature $\{+---\}$,
so that in the limit the symmetry of section is the desired group $SO(1,3)\ltimes T^4$.

Let us find an explicit solution of the equation \eqref{p8.2}
(with $y^{4}$ from \eqref{p8})
with respect to the coordinate $t$:
\disn{p8.4}{
t=\frac{1}{2}\ls -\bar y^{4}\pm \sqrt{{\bar y{}^4}^2+R^2 - x^\m x^\n\eta_{\m\n}}\rs.
\nom}
Each of the roots must be analysed separately. We will discuss the case of choosing the minus sign, and it is easy to show that the alternative choice leads to the same answer.
When choosing the minus sign, at $\bar y^4\to\infty$ the leading approximation is $t=-\bar y^4$.
For the convenience of the analysis we pass to the light-like coordinates in the ambient space by introducing the coordinates
\disn{p9}{
y^+ = \frac{y^5+y^4}{2},\qquad
y^- = \frac{y^5-y^4}{2}
\nom}
instead of the coordinates $y^{4}$ and $y^{5}$.
Then the embedding \eqref{p8} can be rewritten in a simpler form
\disn{p10}{
y^{\m} =x^\m,\qquad
y^+ =\frac{1}{4t} (R^2 - x^\m x^\n\eta_{\m\n}),\qquad
y^- =t.
\nom}
Substituting the solution \eqref{p8.2} found in the leading approximation for $\bar y^4\to\infty$ we obtain the embedding
\disn{p11}{
y^{\m} =x^\m,\qquad
y^+ =-\frac{1}{4\bar y^4} (R^2 - x^\m x^\n\eta_{\m\n}),\qquad
y^- =-\bar y^4.
\nom}

Now note that among the transformations $SO(2,4)$ from the symmetry group $AdS_5$ there are hyperbolic rotations in the plane $y^4,y^5$ which are reduced to the transformation
\disn{p12}{
y^+\to k y^+,\qquad
y^-\to \frac{1}{k} y^+
\nom}
in terms of light-like coordinates $y^+,y^-$. Here $k$ is an arbitrary transformation parameter. Making such a transformation with $k=\bar y^4/R$
(note that the parameter $k$ is dimensionless, so we used the existing dimensional quantity $R$ for dimensionlessness)
we can exclude the infinite value $\bar y^4$ from the embedding function \eqref{p11}.
As a result, after an insignificant general translation of the surface we finally obtain the embedding function as
\disn{p13}{
y^{\m} =x^\m,\qquad
y^+ =\frac{1}{4R} x^\m x^\n\eta_{\m\n},\qquad
y^- =0.
\nom}
By construction, it is symmetric with respect to the Poincare group $SO(1,3)\ltimes T^4$.
The fact that the resulting surface has the induced Minkowski metric is
easy to check directly by the formula \eqref{first_formula}.

Symmetric embedding \eqref{p13} of the Minkowski metric constructed by considering a section, the use of which is justified by some transition to the limit, can be obtained directly as a result of using the \cite{statja27} method which was mentioned at the beginning of the \ref{razdtreb} section.
This approach was implemented by A.~Trukhin in his bachelor's thesis.

Unfortunately, the symmetric embeddings \eqref{p8a2},\eqref{p13} of flat metrics discussed in this section
are not unfolded.
This is easy to understand if we notice that both embeddings lie in some codimension 1
plains
of the ambient space.
It means that among the ambient space vectors $b_{\m\n}^a$ at $\m\geqslant\n$ there is no linearly independent set of $N-n$ vectors (in both considered cases $N-n=2$), which means that there is no unfoldness at $N < {n(n+3)}/{2}$.
In the next section, a method is proposed that allows one to construct unfolded embeddings of the Minkowski space, but this method does not provide embeddings with the Minkowski space symmetry. The resulting embeddings will be unfolded, but not symmetric.
The problem of constructing a symmetric unfolded embedding of flat metrics is not yet amenable to the solution.
A possible way is to find all symmetric embeddings of the Minkowski metric for a given dimension of the ambient space $N$ by the method already mentioned at the beginning of this section, proposed in \cite{statja27}, and then check unfoldness by direct computation. However, for sufficiently large $N$ (for example, for the interesting case $N=10$) this problem is very difficult and its solution is beyond the scope of this work.

In connection with the resulting embedding \eqref {p13} it should be noted that it can be easily generalized to the embedding
\disn{p14}{
y^{\m} =x^\m,\qquad
y^+ =B(x^\m),\qquad
y^- =0
\nom}
with a completely arbitrary function $B(x^\m)$.
It is easy to check that this embedding is also an embedding of the Minkowski space metric; however, it does not inherit its symmetry, and, like \eqref{p13}, it is not unfolded.

\section{Explicit unfolded embeddings of flat metrics}\label{razdrazv}
We will look for embeddings of flat metrics (Euclidean and pseudo-Euclidean) which are unfolded by the definition given in section \ref{razdtreb}.

\subsection{Using $q$-free embeddings}\label{razdrazv-q}
As it was said in section \ref{razdbust}, unfolded embeddings are also $(N-n)$-free embeddings in the range of the ambient space
dimension ${n(n+1)}/{2}\leqslant N \leqslant {n(n+3)}/{2}$, therefore in the first place we will discuss the $q$-free embedding of the Minkowski metric proposed in \cite{bustamante}.
Since the opposite is not true, one has to check whether this embedding is unfolded.

The $q$-free embeddings construction method which was proposed in \cite{bustamante} (authors limit themselves to the case $n=4$) consists in splitting the ambient space into the direct sum of two flat subspaces:
\begin{enumerate}
\item the "base" ambient space of signature $\{-\ldots-\}$;
\item the "extra" space of signature $\{+-\ldots-\}$, i.e. it has one and only one time-like direction.
\end{enumerate}
In the base ambient space \textit{base embedding} is defined, with the embedding function $z^A(x^i)$ depending only on spatial coordinates and satisfying two requirements:
\begin{itemize}
  \item the corresponding induced metric is the flat Euclidean metric;
  \item the set of the nine vectors $\{ \partial _i z^A, \dd_j\dd_k z^A \} $ is linearly independent (which means that the base embedding is unfolded).
\end{itemize}
The authors of \cite{bustamante} propose the 11-dimensional base embedding, thus presenting an 11-di\-men\-si\-onal unfolded embedding of the flat 3-dimensional Euclidean metric.

As for the extra space, it is proposed to construct an extra embedding $w^L(x^\mu)$. In contrast to a base embedding, an extra embedding depends on the time variable $x^0$.
Consequently, the resulting induced metric of the 4-dimensional surface decomposes into
\begin{equation}
g_{\m\n}=\eta _{LM}(\partial _\mu w^L)(\partial _\nu w^M)-\de_\m^i\de_\n^i.
%ds^2 = \eta _{ab} (\partial _\mu w^a)(\partial _\nu w^b)dx^\mu dx^\nu - \delta _{ij} dx^i dx^j .
\end{equation}
In the construction of Minkowski metric embedding, the extra space is considered to be one-dimensional and the extra embedding has the simple form
\disn{p14b1}{
w=x^0.
\nom}
The resulting 12-dimensional embedding is a 6-free embedding, but it is not unfolded, since the rank of the matrix constructed from $b_{\m\n}^a$ (in the manner described after \eqref{p3}) is 6, while the maximum possible rank is $11+1-4=8$.

The situation can be improved by taking a $9$-dimensional base embedding instead of an $11$-dimensional one.
Then the dimension of the transverse space, which in this case defines the maximal possible rank of the matrix constructed from $b_{\m\n}^a$
would equal $9+1-4=6$, thus, the rank would be the highest possible, and the embedding would be unfolded.
We will implement this idea in subsection \ref{razdrazv-mink}.

\subsection{Sequential deformation method}\label{razdrazv-izgib}
Consider the following method of  unfolded embeddings construction for flat metrics of arbitrary dimension $n$:
\begin{enumerate}
\item we start from a simple $(2n)$-dimensional embedding in the form of the direct product of $n$ circles
(pseudoeuclidean circles);
note that it will not be unfolded;
\item we construct a trivial (in the form of a multidimensional plane) $N$-dimensional
(with $N>2n$)
embedding for the resulting $(2n)$-dimensional flat space;
\item on this $(2n)$-dimensional plane we choose $N-2n$ mutually orthogonal straight lines;
\item using $N-2n$ transverse directions, we sequentially deform each of the $N-2n$ mentioned lines into a circle; such a deformation of the $2n$-dimensional plane is an isometric one;
\item we check whether the obtained embedding is unfolded.
\end{enumerate}

\subsection{9-dimensional unfolded (and free) embedding of the Euclidean 3-dimensional metric}\label{razdrazv-3v9}
First, we will use the proposed method to construct an explicit unfolded embedding of the 3-dimensional flat Euclidean metric into the Euclidean 9-dimensional space, so $n=3$ and $N=9$.
To simplify the formulas, we will not introduce dimensional coefficients. Thus, dimensional considerations will be inapplicable to the coordinates.
In accordance with the first step of subsection \ref{razdrazv-izgib}, we start from the embedding
\begin{equation}\label{p14a1}
\begin{array}{lll}
\begin{array}{l}
z^1(x^i) = \sin x^1 , \\
z^2(x^i) = \cos x^1 ,
\end{array} &
\begin{array}{l}
z^3(x^i) = \sin x^2 , \\
z^4(x^i) = \cos x^2 ,
\end{array} &
\begin{array}{l}
z^5(x^i) = \sin x^3 , \\
z^6(x^i) = \cos x^3.
\end{array}
\end{array}
\end{equation}
It is easy to see that locally the metric induced by this embedding is the metric of the Euclidean space (see the Conclusion for a discussion of its global structure).
In order to choose 3
straight
lines in a nontrivial way in accordance with step 3 of subsection \ref{razdrazv-izgib}, we perform a $SO(6)$ rotation
\begin{equation}\label{p14a2}
{z'} ^A (x^i) = O^{AB} z^B (x^i),
\end{equation}
(where  $A,B = 1, \ldots, 6$), then we simply choose 3 coordinate axes $z'^4,z'^5,z'^6$.
It is clear that the induced metric will not change after such a transformation.
One can, for example, choose the matrix
\disn{p14.1}{
O=\exp\ls
\begin{array}{cccccc}
0 & 1 & 0 & -1 & 0 & 0 \\
-1 & 0 & 0 & 0 & 0 & 0 \\
0 & 0 & 0 & -1 & 0 & -1 \\
1 & 0 & 1 & 0 & 0 & 0 \\
0 & 0 & 0 & 0 & 0 & 1 \\
0 & 0 & 1 & 0 & -1 & 0 \\
\end{array}
\rs
\nom}
as an orthogonal matrix $O$, since the matrix exponent of an antisymmetric matrix is an orthogonal matrix.
Depending on the choice of the matrix $O$, the embedding obtained by this method can be either unfolded or not.
For example, one can check that if we take the identity matrix as the matrix $O$, then even after step 4 of subsection \ref{razdrazv-izgib} the resulting embedding will not be unfolded.

Now, in accordance with step 2 of subsection \ref{razdrazv-izgib}, we add 3 trivial directions of the ambient space by introducing the 9-dimensional embedding function $\tilde y^a$:
\disn{p15a}{
\tilde y^A(x^i)=z'^A(x^i),\qquad
\tilde y^7(x^i)=\tilde y^8(x^i)=\tilde y^9(x^i)=0.
\nom}
Obviously, this 9-dimensional embedding is not unfolded.
And finally, in accordance with step 4 of subsection \ref{razdrazv-izgib},
we make an isometric deformation of the $6$-dimensional plane $\tilde y^1,\ldots,\tilde y^6$ in three transverse directions $\tilde y^7,\tilde y^8,\tilde y^9$ so that three coordinate lines $\tilde y^4,\tilde y^5,\tilde y^6$ turn into circles:
\disn{p15}{
\begin{array}{ll}
y^1(x^i)=z'^1(x^i),\qquad &
 y^4(x^i)=\cos z'^4(x^i),\\
y^2(x^i)=z'^2(x^i), &
 y^5(x^i)=\sin z'^4(x^i),\\
y^3(x^i)=z'^3(x^i), &
 y^6(x^i)=\cos z'^5(x^i),\\
& y^7(x^i)=\sin z'^5(x^i),\\
& y^8(x^i)=\cos z'^6(x^i),\\
& y^9(x^i)=\sin z'^6(x^i).\\
\end{array}
\nom}
Since after such a deformation the metric of the 6-dimensional plane does not change, the metric of the 3-surface defined by this embedding also remains flat.
This can also be checked directly by substituting the
embedding \eqref{p15} into the induced metric formula \eqref{first_formula} considering \eqref{p14a1}-\eqref{p14.1}.
Further, following step 5 of subsection \ref{razdrazv-izgib}, it is necessary to check whether the obtained embedding is unfolded.

According to the definition (see section \ref{razdtreb}), an embedding is unfolded if the rank of the matrix composed from the second fundamental form $b_{ik}^a$ in the way described below \eqref{p3} is the maximum possible one.
In the given case this matrix is a square $6\times6$ matrix (the pair of symmetric indices $ik$ corresponds to the multi-index are running through 6 values, and there are the same number of transverse directions).
Therefore, for this matrix the unfoldedness condition coincides with the non-singularity condition.
Direct calculation shows that for the embedding \eqref{p15} this matrix is indeed non-singular, thus the constructed embedding of the 3-dimensional flat Euclidean metric into the Euclidean 9-dimensional space is unfolded.
It is also \emph{free} due to the fulfillment of the condition \eqref{p6.1}.

\subsection{10-dimensional unfolded embedding of the Minkowski metric}\label{razdrazv-mink}
The embedding obtained in the previous subsection can be easily applied to construct an unfolded embedding of the Minkowski metric. In terms of \cite{bustamante} for this one need to use it as a base embedding (see subsection \ref{razdrazv-q}) with a simple extra embedding \eqref{p14b1}.
As a result, we get the 10-dimensional embedding
\disn{p17}{
\begin{array}{ll}
y^0(x^\m)=x^0, &
 y^4(x^\m)=\cos z'^4(x^i),\\
y^1(x^\m)=z'^1(x^i),\qquad&
 y^5(x^\m)=\sin z'^4(x^i),\\
y^2(x^\m)=z'^2(x^i), &
 y^6(x^\m)=\cos z'^5(x^i),\\
y^3(x^\m)=z'^3(x^i), & y^7(x^i)=\sin z'^5(x^i),\\
& y^8(x^\m)=\cos z'^6(x^i),\\
& y^9(x^\m)=\sin z'^6(x^i)\\
\end{array}
\nom}
into the ambient space of signature $\{+-\ldots-\}$.
Since for this embedding the 3-dimensional components $b_{ik}^a$ of the second fundamental form $b_{\m\n}^a$ coincide
with the ones corresponding to the embedding \eqref{p15}, the rank of the corresponding matrix is equal to the maximum possible
value 6 (the number of transverse directions), which means that the embedding \eqref{p17} is unfolded.
Note that the condition \eqref{p6.1} is not satisfied. Therefore, this embedding is not \emph{free} but only \emph{6-free}.

\subsection{14-dimensional unfolded (and free) embedding of the Minkowski metric}\label{razdrazv-4v14}
Now we will apply the method described in subsection \ref{razdrazv-izgib} to the construction of an explicit unfolded embedding of  Minkowski metric into a 14-dimensional space so that $n=4$ and $N=14$.
Since the requirement \eqref{p6.1} is satisfied, the resulting embedding will also be \emph{free}.

Following the first step of subsection \ref{razdrazv-izgib}, we start from the embedding
\begin{equation}\label{p18}
\begin{array}{ll}
z^0(x^\m) = \sinh x^0,\qquad &
 z^4(x^\m) = \sin x^2,\\
z^1(x^\m) = \cosh x^0,&
 z^5(x^\m) = \cos x^2,\\
z^2(x^\m) = \sin x^1,&
 z^6(x^\m) = \sin x^3 ,\\
z^3(x^\m) = \cos x^1,&
 z^7(x^\m) = \cos x^3
\end{array}
\end{equation}
into the ambient space of signature $\{+-\ldots-\}$.
Note that this embedding is the direct product of the
pseudoeuclidean circle
${z^1}^2-{z^0}^2=1$ and three circles.
It is easy to check that the corresponding induced metric is the Minkowski metric.
Next, we will proceed in the same way as in \ref{razdrazv-3v9}.

In order to select 6 straight lines in a nontrivial way in accordance with the third step of subsection \ref{razdrazv-izgib}, we make a $SO(1,7)$ rotation
\begin{equation}\label{p19}
{z'} ^A (x^\m) = \La^A{}_B z^B (x^\m),
\end{equation}
(here $A,B = 0, \ldots, 7$), for example, using the orthogonal matrix
\begin{equation}
\Lambda =\exp
\begin{pmatrix}
0 & 0 & 0 & 0 & 0 & 0 & 0 & 0 \\
0 & 0 & 1 & 0 & 0 & -1 & 0 & 0\\
0 & -1 & 0 & -1 & 0 & 0 & 0 & 0 \\
0 & 0 & 1 & 0 & -1 & 0 & 0 & 0 \\
0 & 0 & 0 & 1 & 0 & 0 & 1 & 0 \\
0 & 1 & 0 & 0 & 0 & 0 & 0 & 1\\
0 & 0 & 0 & 0 & -1 & 0 & 0 & 0 \\
0 & 0 & 0 & 0 & 0 & -1 & 0 & 0 \\
\end{pmatrix},
\end{equation}
then we select 6 coordinate axes $z'^3,\ldots z'^8$.

In accordance with the second step of subsection \ref{razdrazv-izgib}, we add 6 trivial space-like directions of the ambient space, introducing the 14-dimensional embedding function $\tilde y^a$ in which $\tilde y^A(x^i)=z'^A(x^i)$ while the other components with $a=8,\ldots,13$ are equal to zero.
Further, in accordance with the fourth step of subsection \ref{razdrazv-izgib}, we make an isometric deformation of the $8$-dimensional plane corresponding to the directions $\tilde y^0,\ldots,\tilde y^7$ in six transverse directions $\tilde y^8,\ldots,\tilde y^{13}$ with 6 coordinate lines $\tilde y^2,\ldots,\tilde y^7$ turning into circles:
\disn{p20}{
\begin{array}{ll}
y^0(x^\m)=z'^0(x^\m), &
 y^6(x^\m)=\sin z'^4(x^\m),\\
y^1(x^\m)=z'^1(x^\m), &
 y^7(x^\m)=\cos z'^4(x^\m),\\
y^2(x^\m)=\sin z'^2(x^\m),\qquad &
 y^8(x^\m)=\sin z'^5(x^\m),\\
y^3(x^\m)=\cos z'^2(x^\m), &
 y^9(x^\m)=\cos z'^5(x^\m),\\
y^4(x^\m)=\sin z'^3(x^\m), &
 y^{10}(x^\m)=\sin z'^6(x^\m),\\
y^5(x^\m)=\cos z'^3(x^\m),
& y^{11}(x^\m)=\cos z'^6(x^\m),\\
& y^{12}(x^\m)=\sin z'^7(x^\m),\\
& y^{13}(x^\m)=\cos z'^7(x^\m).\\
\end{array}
\nom}

By construction, for this embedding the induced metric \eqref{first_formula} will locally coincide with the Minkowski metric (see the Conclusion for a discussion of its global structure).
One can check that its second fundamental form $b_{\m\n}^a$, interpreted as a $10\times10$ matrix (see the explanation below the formula \eqref{p3}), turns out to be non-degenerate, so this embedding is unfolded.

\section{Conclusion}
Isometric embeddings of flat metrics can be nontrivial, i.e. different from a plane in the ambient space.
Such nontrivial embeddings are of interest from the point of view of describing gravity within the framework of the embedding theory.
We have discussed some possible additional requirements that may apply when searching for these embeddings.
There is an obvious possibility to require the constructed surface to have the symmetry of the original metric.
But there is also an alternative requirement that turns out to be crucial from the point of view of the RT equations \eqref{rt_equations} analysis (in particular, when one tries to linearize them).
%which are the equations of motion for the embedding gravity
This is the requirement of unfoldness.
We call an isometric embedding of a given metric unfolded if the corresponding surface locally occupies a subspace of the maximum possible dimension almost everywhere.
The introduced concept of unfoldness is closely related to the free embedding concept discussed in the work \cite{bustamante}.

We have discussed symmetric embeddings of flat Euclidean three-dimensional space and Minkowski space.
We have proposed the method of sequential deformation of the surface to construct unfolded (but not symmetric) embeddings,
Using this method, we succeeded to construct an unfolded embedding (\ref{p15}) of the metric $\mathds{R}^{3}$ into the ambient Euclidean space $\mathds{R}^{9}$, as well as the unfolded embedding (\ref{p20}) of the Minkowski metric $\mathds{R}^{1,3}$ into the ambient multidimensional Minkowski space $\mathds{R}^{1,13}$.
Based on the embedding (\ref{p15}) an unfolded embedding (\ref{p17}) of the Minkowski metric into $\mathds{R} ^{1,9}$ was also obtained.
Note that the proposed method of sequential deformation
can also be used to build new multidimensional embeddings based on already known embeddings with a small value of the embedding class. For example, this method can be applied to the known (see \cite{statja27}) 6-dimensional black hole embeddings. However, whether it is possible to obtain unfolded embeddings of black holes in this way is the question that requires a separate study.

It should be noted that for all proposed embeddings (\ref{p15}),(\ref{p17}),(\ref{p20}) the embedding functions are periodic in spatial coordinates by construction.
This means that the surfaces defined by these embeddings are compact, thus their topology differs from the topology of the original metrics.
From a physical point of view, within the framework of embedding gravity such a property of the background embedding function $\zero{y}^a(x)$ does not agree well with observations.
Indeed, even if the universe is compact in spatial directions (as, for example, within the framework of the closed FRW model), then the period has to be very large, no less than the size of the visible part of the universe $L$.
We can provide such a value of the period by introducing the value $L$ into the formulas \eqref{p14a1},\eqref{p18}.
If we replace $x^1$ with $x^1/L$ and so on, then the coordinates become dimensional values, and the period becomes equal to $2\pi L$.
But then the part of the surface with a size of the order of the galaxy diameter will be practically indistinguishable from the plane, i.e. the unfoldness of the embedding will no longer be visible.
This corresponds to the fact that the second fundamental form $b^a_{\m\n}$ will be of order $1/L$, i.e. it will be very small.
Therefore, it is necessary to get rid of the periodic property of the embedding functions.

It turns out that for unfolded embeddings for $N= n(n+3)/2$ this can be easily done by an infinitesimal non-periodic isometric deformation of the surface (a deformation that does not change the metric locally).
For this deformation, the left side of the equation \eqref{urav_bustamante} is zero.
We can take an arbitrary non-periodic function $\xi^\m(x)$ which defines the longitudinal part of the deformation, and from the vanishing of the right-hand side of \eqref{urav_bustamante} we find the transverse part of the deformation $\delta y^a _{\perp}$,
which will also be non-periodic.
For $N= n(n+3)/2$ for an unfolded embedding this can always be done easily by inverting the value $b^a_{\m\n}$ interpreted as a matrix in the sense described below the formula \eqref{p3}.
This matrix will be a square matrix of size $n(n+1)/2$; its non-singularity follows from the definition of an unfolded embedding.
The described  periodicity elimination of the embedding is easiest to imagine by noting that it occurs in the same way as in the transition from a circle to a spiral.
This procedure can be directly applied to unfolded embeddings (\ref{p15}) and (\ref{p20}) for which $N= n(n+3)/2$ in order to break the periodicity of them.
A non-periodic analogue of the embedding (\ref{p17}) can be obtained by a procedure similar to the one described in the \ref{razdrazv-mink} section, but taking the deformed embedding \eqref{p15} as the base embedding the periodicity of which has already been eliminated.
Note that the resulting non-periodic embeddings turn out to be global embeddings for the considered non-compact flat spaces.

The resulting embeddings and their deformations with broken periodicity can be used both in the analysis of the properties of extra matter in the non-relativistic
limit of
embedding gravity \cite{statja68} and as a background in the study of more general solutions of the RT equations \eqref{rt_equations} in the limit of a weak gravitational field \eqref{pp1}.

{\bf Acknowledgements.}
The authors thank A.~Trukhin who obtained the embedding \eqref{p13} in an alternative way.
The work is supported by RFBR Grant No.~20-01-00081.

%\newcommand{\eprint}[1]{\href{http://arxiv.org/abs/#1}{\texttt{#1}}}
%\bibliographystyle{../../my3}
%\bibliography{../../paston-grav-e}
%\end{document}

\end{document}